\documentclass[pdflatex,sn-mathphys,iicol]{sn-jnl}

\usepackage{amsmath}
\usepackage{graphicx}
\usepackage{float}
\usepackage{xcolor}

\def\po{p_1}
\def\pz{p_0}
\def\ppo{p^{\prime}_1}
\def\ppz{p^{\prime}_0}

\jyear{2021}%

\theoremstyle{thmstyleone}%
\theoremstyle{thmstyletwo}%

\theoremstyle{thmstylethree}%

\begin{document}

\title[Thermodynamics of multiple Maxwell demons]{Thermodynamics of multiple Maxwell demons}


\author*[1]{\fnm{Sandipan} \sur{Dutta}}\email{sandipan.dutta@pilani.bits-pilani.ac.in}

\affil*[1]{\orgdiv{Department of Physics}, \orgname{Birla Institute of Technology and Science}, \orgaddress{\city{Pilani}, \postcode{333031}, \state{Rajasthan}, \country{India}}}


\abstract{In many assembly line processes like metabolic and signaling networks in biological systems, the products of the first enzyme is the reactant for the next enzyme in the network. Working of multiple machines leads to efficient utilization of resources. Motivated by this, we investigate if multiple Maxwell demons lead to more efficient information processing. We study the phase space of multiple demons acting on an information tape based on the model of Mandal and Jarzynski \cite{mandal2012work, mandal2013maxwell}. Their model is analytically solvable and the phase space of the device has three regions: engine, where work is delivered by writing information to the tape, erasure, where work is performed on the device to erase information on the tape, and dud, when work is performed and at the same time the information is written to the tape. For identical demons, we find that the erasure region increases at the expense of the dud region while the information engine region does not change appreciably. The efficiency of the multiple demon device increases with the number of demons in the device and saturates to the equilibrium (maximum) efficiency even at short cycle times for very large number of demons. By investigating a device with nonidentical demons acting on a tape, we identify the demon parameters that control the different regions of the phase space. Our model is well suited to study information processing in assembly line systems.}




\maketitle

\section{Introduction}\label{sec1}

Maxwell demon (MD) emerged out of a thought experiment by James C. Maxwell \cite{maxwell1871theory} that created a paradox in the second law of thermodynamics. In the experiment the demon controls a trap door that separates two chambers of gases both at the same temperature. The fast molecules of the gas are prevented by the demon to cross to the other chamber, only allowing the slow ones to pass through. This creates order reducing the entropy apparently violating the second law of thermodynamics. Szilard \cite{szilard1929entropieverminderung} proposed another realization of the demon that can extract $k_BT\ln 2$ of work using one bit of information from the position of a molecule in a box in contact with a reservoir of temperature $T$. Landauer  \cite{landauer1961irreversibility, PhysRevLett.125.100602, esposito2011second} gave a physical realization of information. He asserted that a minimum amount of energy of $k_BT\ln 2$ is required to erase one bit of information. Bennett \cite{bennett1982thermodynamics, feynman2018feynman} and Penrose  \cite{penrose2005foundations} showed that a mechanical demon can rectify thermal fluctuations to perform work but in the process the information has to be written to a memory register. These seminal works led to the physical realization of information that helped unravel the paradox of the MD. Since then a vast literature that introduced information into thermodynamics has been developed \cite{sagawa2010generalized, parrondo2015thermodynamics,  abreu2011extracting, abreu2012thermodynamics, horowitz2014thermodynamics, seifert2012stochastic, cao2015thermodynamics, barato2014unifying, horowitz2013imitating, cao2009thermodynamics, cao2004feedback, ponmurugan2010generalized, horowitz2011thermodynamic, kundu2012nonequilibrium, barato2013autonomous, kawai2007dissipation, horowitz2010nonequilibrium} . MDs based on measurement and feedback \cite{toyabe2010experimental, koski2014experimental, paneru2018lossless, paneru2020efficiency, paneru2020reaching} and Landauer principle \cite{PhysRevLett.117.200601} have been experimentally realized and have found applications in the signal transduction in biological network \cite{tu2008nonequilibrium, sartoriplos, ito2015maxwell, andrieux2008nonequilibrium}, in quantum information processing \cite{chuan2012quantum, jacobs2012quantum, quan2006maxwell} and in the artificial nanoscale devices \cite{serreli2007molecular, strasberg2013thermodynamics}. 

\begin{figure}[!htb]
\includegraphics[width=0.48\textwidth]{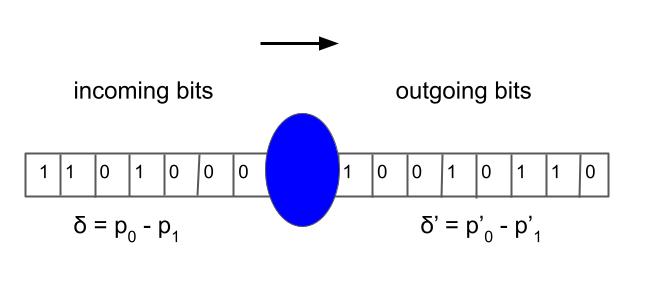}
\caption{The schematic diagram of a demon processing an information tape. The demon interacts with the tape for a time $t$ which results in transitions between the internal states $A$, $B$ and $C$ of the demon as in Fig \ref{fig:transition1}. The probability of excess 0s in the incoming bits compared to 1 is called the \textit{excess parameter} $\delta$, which changes after the interaction to $\delta^{\prime}$. }
\label{fig:singledemon}
\end{figure}

Many systems often involve not one but multiple demons working in an "assembly line" where the products of the first demon is processed by the second demon and so on. Examples include the metabolic and signaling networks in biology. It is important to study the thermodynamics of information processing in such systems, in particular how the thermodynamics of a multiple demon system is different from a single demon or if a multiple demon system processes information more efficiently than a single demon.  Here we model the MDs based on an exactly solvable model of MD by Mandal and Jarzynski \cite{mandal2012work, mandal2013maxwell} where the demon, having discrete states, interacts with an information tape of $0$s and $1$s in presence of a thermal reservoir of temperature $T$ as shown in Fig \ref{fig:singledemon}. This interaction induces transition between the states of the demon and simultaneously flipping the bits in the tape resulting in lifting or lowering of weights, thereby performing work. All the three regimes of the MD can be realized through the toy model - engine, erasure and dud. As an engine, work can be extracted by exploiting the thermal fluctuations and writing information to the memory tape. The MD acts as an erasure when an external work is performed to erasure the information from the tape. Finally in the dud region, the disorder of the tape increases and simultaneously work has to be performed on the demon, thus not accomplishing nothing useful. The entropy of the tape increases (decreases) for the engine (erasure) while the work is extracted (performed). Mandal and Jarzynski model is a realization of MD without any explicit measurement and feedback. In this work we consider multiple demons each in steady state acting on a single tape and investigate how this tape reaches steady state due the repeated application of the demons. We find that when all demons are identical then the erasure region of the phase space increases with the number of demons while the information engine region remains constant. To identify the parameters which control the different phase space regions, we study nonidentical demons with different transition rate parameters and interaction time with the tape. We find that while the erasure region increases with increasing interaction time, the behavior of the engine region with the transition rate parameter of one of the demons is non-monotonic.

\begin{figure}[!htb]
\includegraphics[width=0.48\textwidth]{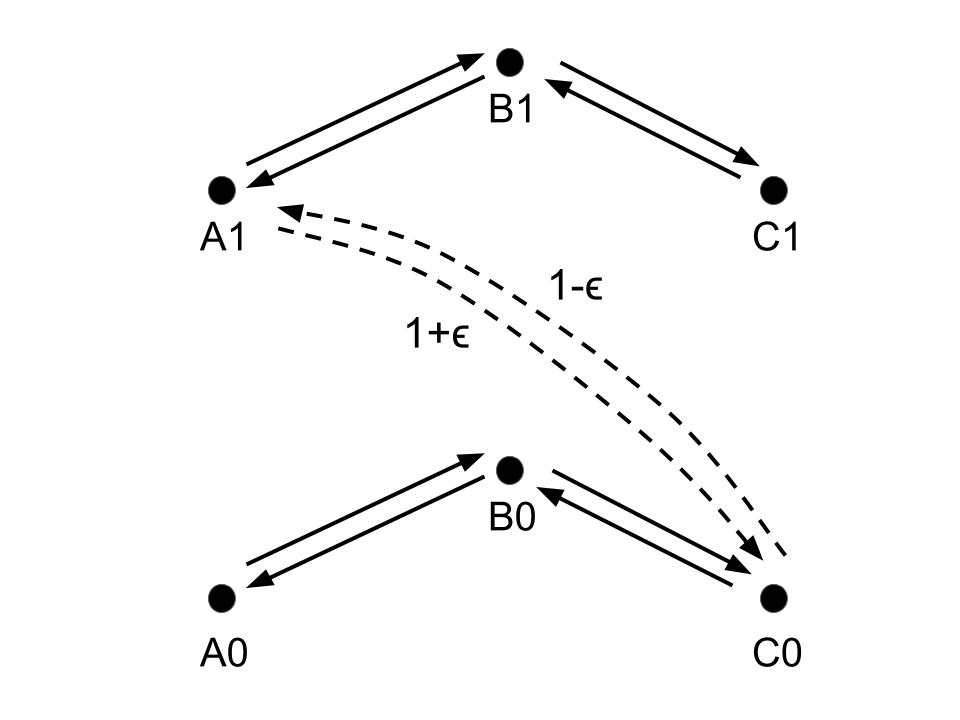}
\caption{The allowed transition in the demon-bit system with the dashed lines showing the coupling between the demon and bit. The rates are unity for all the solid arrows. The \textit{bias parameter} $\epsilon$ introduces a bias to the transition $A1\rightarrow C0$ over $C0\rightarrow A1$. }
\label{fig:transition1}
\end{figure}

We start with a three-state model of the identical MDs working on a bit stream and look at its phase space. Then we introduce a system of multiple identical MDs working independently on a single tape and look at the thermodynamics and the efficiency of the composite demon system. In order to investigate how the thermodynamics of the multiple MDs depends on the particular model of the demon, we also study the thermodynamics of MDs with two states \cite{boyd2016identifying}. Finally we investigate how the differences of the rate parameters and the interaction time with the tape affects the thermodynamics of a non-identical demon system.

\section{Model of a Demon}\label{sec2}

Based on Ref \cite{mandal2012work}, we consider a three state model of a Maxwell Demon, $A$, $B$ and $C$  interacting with a tape (a sequence of bits) and a thermal reservoir at temperature $T$ as shown in Fig \ref{fig:singledemon}. This interaction, where each bit from the tape interacts for a time $t$ with the demon, induces transitions from $C$ to $A$ ($A$ to $C$) if the bit simultaneously flips from 0 to 1 (1 to 0). The composite states of the demon-tape system are $A0$, $B0$, $C0$, $A1$, $B1$ and $C1$. After interacting with the system, the bit moves forward, and a new bit comes to interact with the tape. This new incoming bit determines the initial state of the next interaction cycle, with the system starting from state A for bit 1 and from state C for bit 0. The transition rate matrix for the demon-bit system corresponding to Fig \ref{fig:transition1} is
\begin{equation} R =
\begin{bmatrix}
-1 & 1 & 0 & 0 & 0 & 0 \\
1 & -2 & 1 & 0 & 0 & 0 \\
0 & 1 & -2+\epsilon & 1+\epsilon & 0 & 0 \\
0 & 0 & 1-\epsilon & -2-\epsilon & 1 & 0 \\
0 & 0 & 0 & 1 & -2 & 1 \\
0 & 0 & 0 & 0 & 1 & -1 \\
\end{bmatrix}.
\end{equation}
All the transition rates $R_{i,j}$ are one except $C0\rightarrow A1$, $R_{A1,C0} = 1-\epsilon$ and $A1\rightarrow C0$, $R_{A1,C0} = 1+\epsilon$ respectively. $\epsilon$ is the \textit{bias parameter} as it biases the system towards certain states that help rectify the thermal fluctuations to extract work. The bias is introduced in the transition rates though a thermodynamic energy $E$ through $R_{A1,C0}/R_{C0,A1} = e^{-E/k_BT}$, which implies $\tanh(E/2k_BT) = \epsilon$. For example, in Ref \cite{mandal2012work} the demon lifts (drops) a mass $m$ through a height $h$ resulting in a thermodynamic energy $E = \pm mgh$. During the transition $C0\rightarrow A1$ heat is absorbed from the reservoir to lift the mass and for the transition $A1\rightarrow C0$, the energy $E$ is released into the reservoir. For the rest of the discussions the work and the energies are made dimensionless by scaling them with respect to the thermal energy $k_BT$. 

For every transition A1 to C0 (C0 to A1), an energy $E$ is extracted (performed). The incoming bits are statistically independent with a probability $\mathbf{p} = (\pz, \po)$, where $\pz$ is the probability of the system to be in state 0, and $\po$ to be in state 1. It is easier to quantify the incoming probabilities by the excess of 0 bits in the incoming stream, the \textit{excess parameter} $\delta = \pz-\po$. Similarly the probability of the outgoing bit in periodic steady state is $\mathbf{p}^{\prime} = (\ppz, \ppo)$, with the probability of bit 1 is $\ppo$ and bit 0 is $\ppz$. 

The transition matrix of the composite system is, $T_p = P_D\exp(Rt)M_p$, where $P_D = (I, I)$ and $M_p = (\pz I, \po I)^{T}$. $I$ is the $3\times 3$ identity matrix. If $\mathbf{P}_p$ is the periodic steady state of the demon, then it satisfies $T_p\mathbf{P}_p = \mathbf{P}_p$. The demon state $\mathbf{P}_p$ and the matrix $T_p$ depends on the incoming bits state $\mathbf{p}$. The steady state probability distribution of the outgoing bits are obtained from the steady state of the demon by
\begin{equation}
\mathbf{p}^{\prime} = P_B\exp(Rt)M_p\mathbf{P}_p,\hspace{5 mm} 
P_B = \begin{pmatrix}
1 & 1 & 1 & 0 & 0 & 0  \\
0 & 0 & 0 & 1 & 1 & 1  \\
\end{pmatrix}.
\end{equation}
The average work extracted for each cycle is
\begin{equation}
W =  k_BT(\ppo - \po)\ln((1+\epsilon)/(1-\epsilon)),
\end{equation}
where $\epsilon = \tanh(E/2)$ is the bias parameter. The amount of information of information written onto bits is the difference between the Shannon entropy $S(p_1) = -p_1\ln p_1 - p_0\ln p_0$ of the outgoing and incoming bits \cite{thomas2006elements}
\begin{equation}
\triangle S =  S(\ppo) -S(\po).
\end{equation}

\begin{figure}[!htb]
\includegraphics[width=0.48\textwidth]{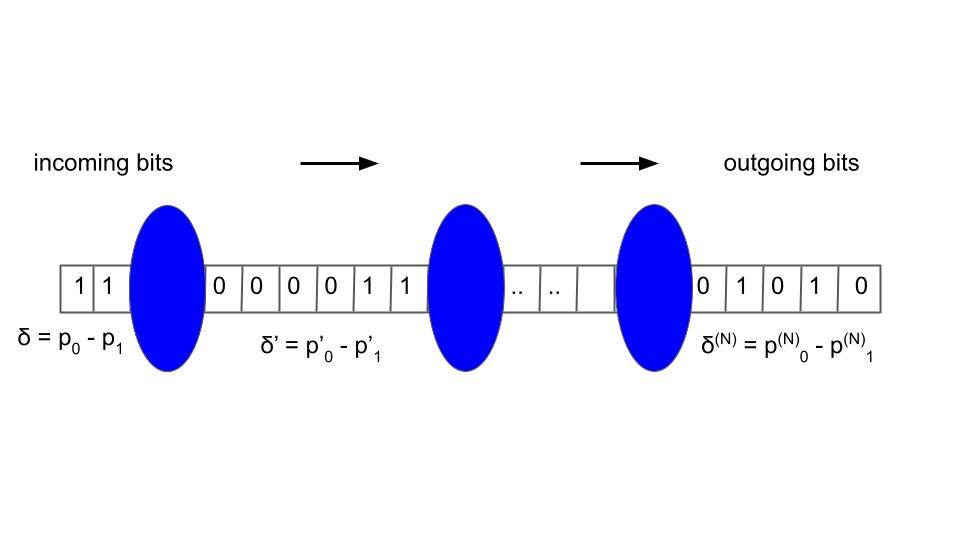}
\caption{The schematic diagram of N demon processing an information tape. The incoming bits of a demon is the outgoing bits of the previous demon.  }
\label{fig:multipledemons}
\end{figure}

\begin{figure*}[!htb]
\includegraphics[width=\textwidth]{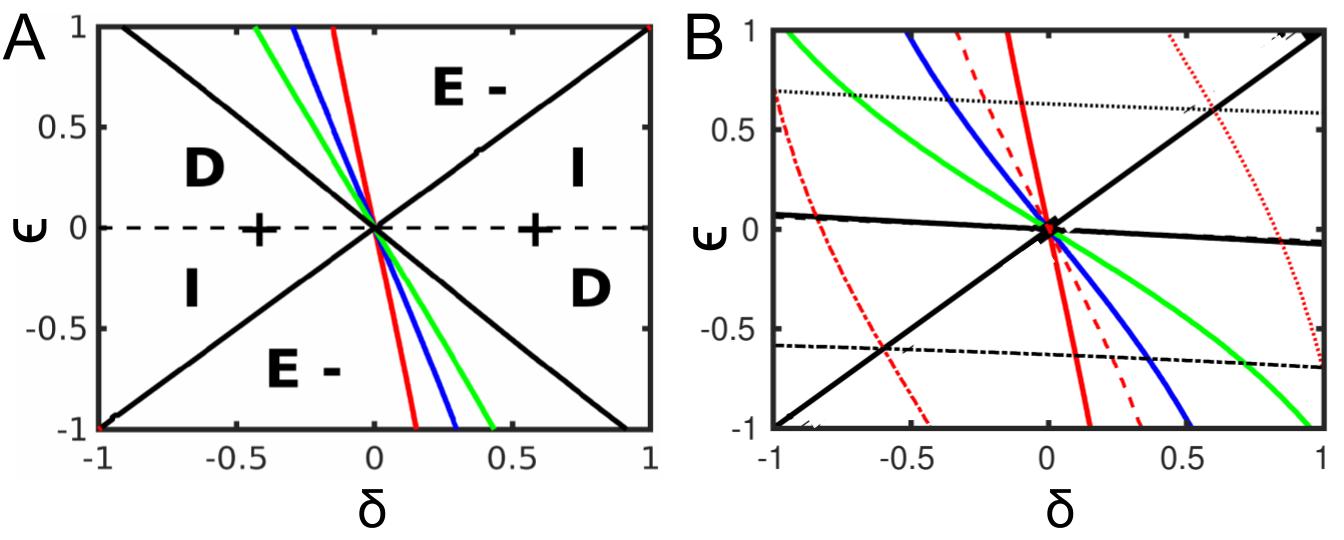}
\caption{(\textbf{A}) The phase diagram of a $N$ identical demon device as a function of the bias parameter $\epsilon$ and the excess parameter $\delta$ of the 0's in the incoming bits of the device for cycle time $t=1$ for different values of $N$. $\triangle S_{0,N}$ vanishes along the lines which demarcate the regions erasure \textbf{E}, engine \textbf{I} and dud \textbf{D}. The red, blue, green and black corresponds to $N=1$, $N=2$, $N=3$ and $N=10$ demon device. The \textbf{E} increases with $N$, while the dud region \textbf{D} decreases and the engine region \textbf{I} remains unchanged. (\textbf{B}) The solid  red, blue, green and black curves show the inversion of the bit statistics, where the probabilities of the incoming and outgoing bits are flipped, for the first, second, third and last demon of a $N = 10$ demon device. The dotted, the dashed and the dashed dotted curves depicts the outgoing bit $1$ probability $0.2$, $0.5$ and $0.8$ respectively for the first demon (red) and the last demon (black). }
\label{fig:phase}
\end{figure*}

\begin{figure*}[!htb]
\includegraphics[width=\textwidth]{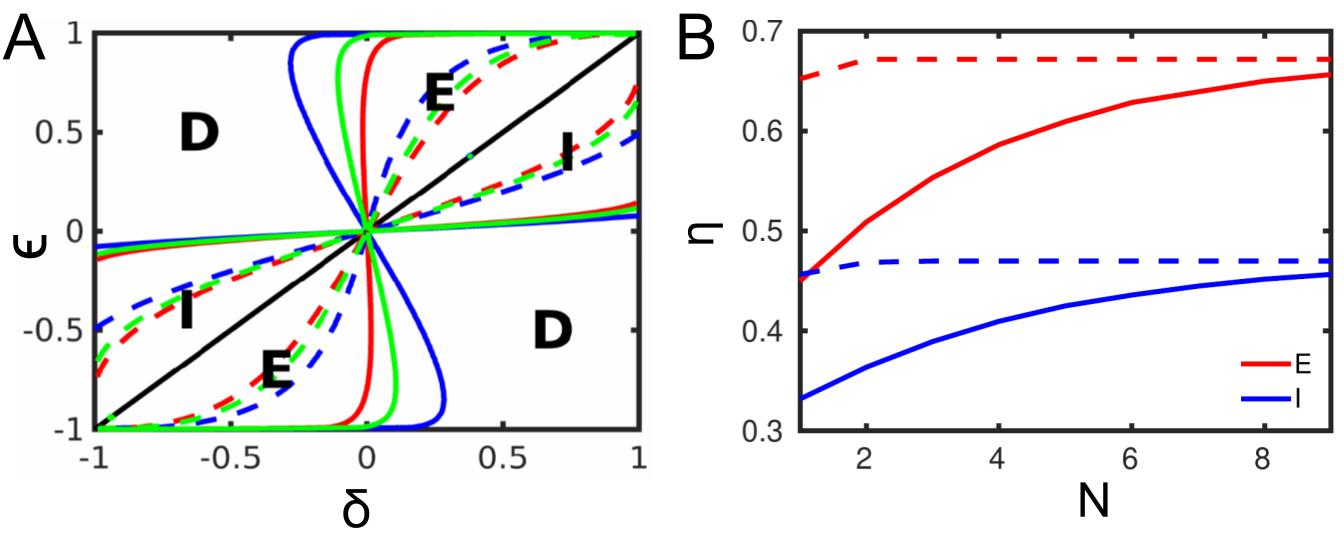}
\caption{(\textbf{A}) The phase diagram of the efficiency $\eta$ of a two demon device for $t = 1$. The solid and dashed curves represent $\eta = 0.1$ and $\eta = 0.5$ for the first demon (red), the second demon (blue) and the device (green). $\eta = \eta_I$ (see text) in the engine regime \textbf{I} and $\eta = \eta_E$ in the erasure regime \textbf{E}. (\textbf{B}) The solid and dashed curves showing the behavior of the efficiency for increasing number of demons for the parameters $(\delta = 0.2, \epsilon = 0.5)$ (red) and ($\delta = 0.6, \epsilon = 0.2$) (blue) at $t = 1$ and $t = 10$ respectively. The parameters are chosen to lie in \textbf{E} (red) and in \textbf{I} (blue). }
\label{fig:efficiency}
\end{figure*}

The phase diagram of the demon for different the bias parameters $\epsilon$ and the excess parameters $\delta$ is shown in red for short cycle time $t=1$ in Figure \ref{fig:phase}-(A). It has three regions: the erasure region \textbf{E} (with $\triangle S < 0$, $W<0$), the dud \textbf{D} (with $\triangle S > 0$ and $W<0$) and the engine \textbf{I} (with $\triangle S > 0$ and $W>0$). The diagonal (black) line $\delta = \epsilon$ on which there is no change in bit statistics: $\ppz = \pz$ and $\ppo = \po$, separates regions \textbf{E} and \textbf{I}. While the red line separates \textbf{D} and \textbf{E} which corresponds to inversion of the bit statistics: $\ppz = \po$ and $\ppo = \pz$. In \textbf{I} the demon can extract work from the reservoir by writing information to the tape, while in \textbf{E} it uses the work to erase the information in the tape and in \textbf{D} the entropy of the tape increases and at the same time work has to be performed on it doing nothing useful.

\section{Multiple identical demons}\label{sec3}

\subsection{Demons with three states}\label{sec3.1}

For a device with $N$ multiple identical demons each in periodic steady state and acting on the same tape, the incoming bits of a demon is the outgoing bits of the previous demon. The outgoing bit probabilities $\mathbf{p}^{\prime}$ of the first demon are calculated from the transition matrix $T_p$ and the steady state distribution of the demon $\mathbf{P}_p$. Since the next demon works on the same tape, this becomes the incoming bit for the second demon. Again the outgoing bit statistics of the second demon $\mathbf{p}^{\prime\prime}$ are obtained from the the transition matrix $T_{p^{\prime}}$ and the steady state distribution of the demon $\mathbf{P}_{p^{\prime}}$, and the process continues. If $\mathbf{p}^{(n)}$ is the outgoing bit probability of the $n^{\text{th}}$ demon, then the change in the total information between the m$^{\text{th}}$ and n$^{\text{th}}$ ($m<n$) demons is $\triangle S_{m,n} = S(p^{(n)}_1)-S(p^{(m)}_1) $ and the work extracted is $W_{m,n} = k_BT(p^{(n)}_1 - p^{(m)}_1)\ln((1+\epsilon)/(1-\epsilon))$.

The phase diagram of $N=1$ demon system (red), $N=2$ demon system (blue), $N=3$ demon system (green) and $N=10$ demon system (black) for cycle times $t=1$  are shown in Figure \ref{fig:phase}-(A). At short cycle times, as the number of demons processing the tape increases, the erasure region \textbf{E} also increases at the expense of the dud region \textbf{D}, while the engine region \textbf{I} remains unchanged. At long cycle times, the bit-demon system gets enough time to interact and relax to equilibrium so having multiple demons does not change the phase space much. This can be understood from the behavior of $W = 0$ and $\triangle S = 0$ which demarcates the \textbf{E}, \textbf{I} and \textbf{D} regions. While $W = 0$ contours do not change with the demons (not shown), the $\triangle S = 0$ regions increase with for the higher order demons as seen in Figure \ref{fig:phase}-(B).   

\begin{figure*}[!htb]
\includegraphics[width=\textwidth]{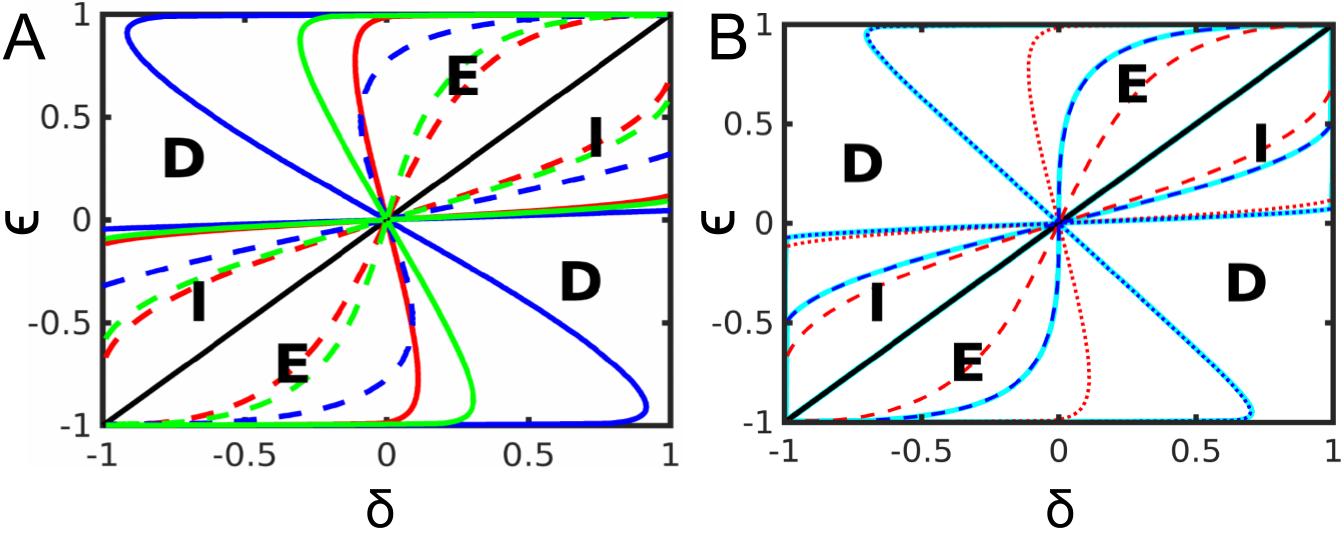}
\caption{ (\textbf{A}) The phase diagram of the efficiency $\eta$ for a two demon device with two states for $t = 1$ has a similar behavior as the device with three states in Fig \ref{fig:efficiency}-(A). (\textbf{B}) Comparison of the constant efficiency curves $\eta = 0.1$ (dotted) and $\eta = 0.5$ (dashed) of a single demon device (red) and $N = 10$ demon device (blue) for $t = 1$  with the maximum efficiency (cyan) obtained from Equation \eqref{eq:efficiency}.}
\label{fig:2states}
\end{figure*}

\begin{figure*}[!htb]
\includegraphics[width=\textwidth]{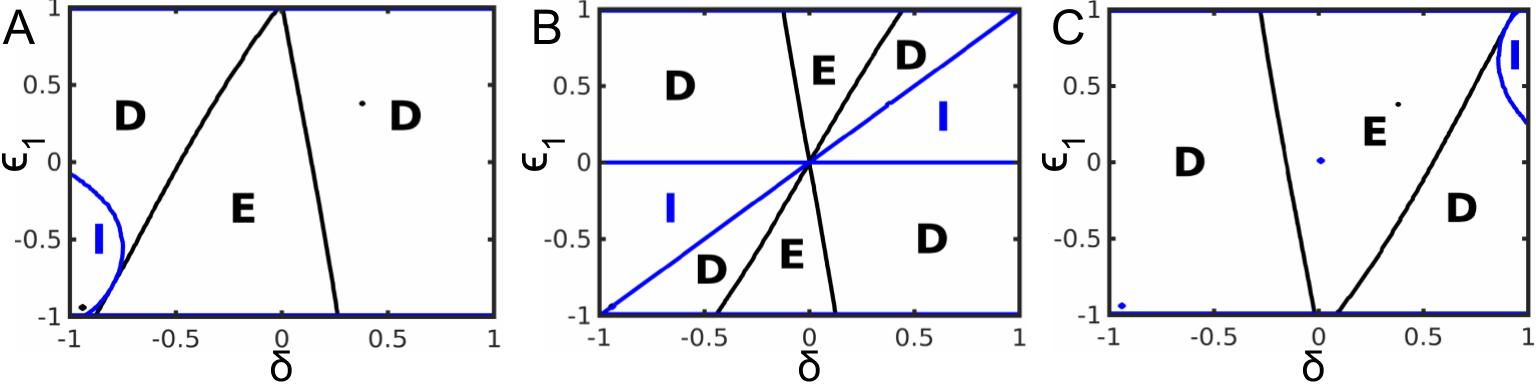}
\caption{The phase diagram of two demons with different bias parameters $\epsilon_1$ and $\epsilon_2$ with fixed (\textbf{A}) $\epsilon_2 = -0.8$ , (\textbf{B}) $\epsilon_2 = 0$ and (\textbf{C}) $\epsilon_2 = 0.8$ respectively for $t = 1$.}
\label{fig:twodemons}
\end{figure*}

We also study the interplay of the energy and information for multiple demons using the efficiency of the device. Since the engine \textbf{I} performs work by utilizing the information of the tape, the efficiency of the device is defined as $\eta_I = W/\triangle S$. The efficiency of an engine with $N$ demons  would be $\eta_I^{(N)} = W_{0,N}/\triangle S_{0,N}$. Similarly the efficiency of the $n^{\text{th}}$ demon in a $N$ demon assembly is $\eta_I^{(n)} = W_{n-1,n}/\triangle S_{n-1,n}$. As an erasure \textbf{E}, work has to be performed on the device to erasure information, hence the efficiency of the device is $\eta_E = \triangle S/W$. Similarly, for a multiple erasure device, the efficiency is $\eta_E^{(N)} = \triangle S_{0,N}/W_{0,N}$. The constant efficiency curves $\eta = 0.1$ and $\eta = 0.5$ of a two demon device in Fig \ref{fig:efficiency} -(A) shows that the efficiency of the second demon is higher than the first demon. To understand this behavior we focus on the erasure region \textbf{E} in the upper half plane where the efficiency increases to the right. The constant efficiency curve $\eta = 0.5$ for the first demon is always on the right of the $\eta = 0.5$ curve for the second demon. Hence all the points on the $\eta= 0.5$ curve of the first demon have higher efficiencies for the second demon. This argument applies to all any constant efficiency curves and also to region \textbf{I}. Thus the second demon has higher efficiency which immediately implies the two demon device has higher efficiency then a single device consisting of the first demon. This finding is more general and in Fig \ref{fig:efficiency} -(B) we see that the efficiency of a $N$ demon device for $t = 1$ increases with increasing number of demons at two given points in phase space and then saturates to a maximum value (Fig \ref{fig:efficiency} -(B)) corresponding to the equilibrium efficiency $t = 10$. Thus a $N$ demon device can have equilibrium maximum efficiency even at short cycle times for sufficiently large $N$. The maximum efficiency of the system as derived in Ref \cite{mandal2012work}) 
\begin{align}
\eta_I^{\text{max}} & = \frac{\delta-\epsilon}{2(S(\delta)-S(\epsilon))}\ln\left(\frac{1+\epsilon}{1-\epsilon}\right), \hspace{1mm} \text{for engine \textbf{I} and,} \\
\eta_E^{\text{max}}& = \frac{2(S(\delta)-S(\epsilon))}{\delta-\epsilon}\frac{1}{\ln\left(\frac{1+\epsilon}{1-\epsilon}\right)}, \hspace{7mm} \text{for erasure \textbf{E}}.
\label{eq:efficiency}
\end{align}
This behavior can be understood from the fact that repeated application of the multiple demons on the same tape for a finite cycle time $t$ is effectively equivalent to a demon with long cycle time. Increasing number of demons quasi-statically drives the system to the maximum efficiency $\eta^{\text{max}}$ attained at equilibrium. More demons with shorter cycle times would be needed to reach $\eta^{\text{max}}$ than the ones with larger cycle times.

\subsection{Demons with two states}\label{sec3.2}

To check the generality of the observations of the previous section, we look at demons having two states $A$ and $B$, which upon interaction with the tape has composite states $A0$, $B0$, $A1$ and $B1$. The transition matrix for this case is
\begin{equation} R =
\begin{bmatrix}
-1 & 1 & 0 & 0 \\
 1 & -2+\epsilon & 1+\epsilon & 0 \\
0 & 1-\epsilon & -2-\epsilon & 1 \\
0 & 0 & 1 & -1 \\
\end{bmatrix}.
\end{equation}
The phase diagram is similar to the three state demon, with a larger erasure region. This shows up in the efficiency contour plot of two demons in Fig \ref{fig:2states}-(A). The maximum efficiency for a demon with two states is the same as for three states in Equation \ref{eq:efficiency}. In Fig \ref{fig:2states}-(B) the efficiency of $N=10$ demon device attains the maximum efficiency even at short cycle time $t = 1$ similar to the behavior of a device with three states in Fig \ref{fig:efficiency}-(B).

\begin{figure}[!htb]
\includegraphics[width=0.48\textwidth]{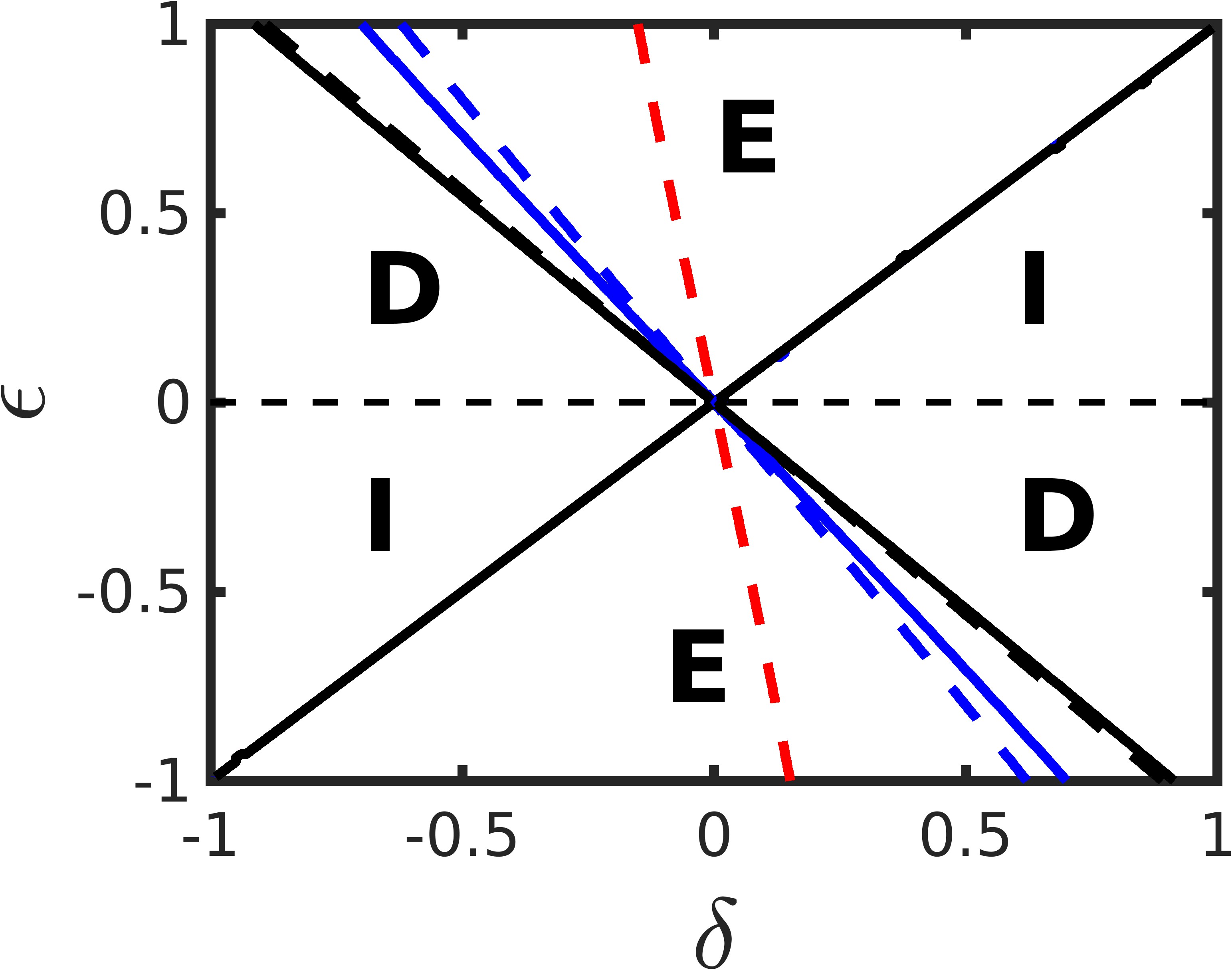}
\caption{The phase diagram of two demon device with different cycle times $t_1$ and $t_2$ but same bias parameters. The dashed lines are $\triangle S = 0$ for single demon device of cycles $t = 1$ (red), $t = 5$ (blue) and $t = 10$ (black). The solid lines are the ones for the two demon device with $t_1 = 1, t_2 = 5$ (blue) and $t_1 = 1, t_2 = 10$ (black).  }
\label{fig:differenttimes}
\end{figure}

\section{Two non-identical Demons}\label{sec3.1}
Until now we have looked at identical demons acting on a tape that is all the demons have same bias parameters and acting on the tape for same amount of time. Here we investigate how a composite system of two demons with different bias parameters behaves. We consider a two demon system with bias parameter of the demon one $\epsilon_1$ and demon two $\epsilon_2$ but with same interaction time $t$. We plot the phase space of the system for $\epsilon_1$ and $\delta$ for three fixed values of $\epsilon_2$ in Fig \ref{fig:twodemons} for short cycle time $t = 1$. We see that the two demon system works mostly as an erasure \textbf{E} with a very small engine regime \textbf{I} for large $\epsilon_2$. \textbf{I} region occurs around $\epsilon_1 \approx \epsilon_2$. \textbf{I} attains maximum when $\epsilon_2 = 0$ where the dud region attains minimum. Similar behavior is observed for long cycle time $t = 10$, however we have larger erasure region. Investigation of two demons with identical bias parameters but acting on the tape for different times show that the the demon which acts for longer time on the tape dominates the behavior of the combined system. Fig \ref{fig:differenttimes} shows that the phase diagram of a two demon device with the first demon having time $t_1 = 1$ and the second demon with $t_2 = 5$ and $t_2 = 10$, looks similar to a single device with $t = 5$ and $t = 10$ respectively.

\section{Conclusion}\label{sec13}

We have studied the effect of multiple Maxwell demons working on a tape. A model of an autonomous information processing device without any explicit measurement and feedback has been used to model the demons based on Ref \cite{mandal2012work, mandal2013maxwell}. For identical demons, we found that the erasure region increases with the number of demons while the dud region decreases while the information engine region remains unchanged. The efficiency can be increased by increasing the number of demons and attains the maximum efficiency even at short cycle times for very large number of demons. We found similar phase diagrams and efficiencies for demons with two states. Investigation of nonidentical demons where demons have different bias parameters and interaction time with the demons show that increasing the time for one of the demons increases the erasure region. However, increasing the bias parameter for one demon keeping the other demon constant in a two demon device, changes the engine region non-monotonically for identical cycle times. This work can be used to study assembly line processes like sensory-receptor systems which involve information processing.

\bmhead{Acknowledgments}

The authors acknowledge Birla Institute of Technology and Science, Pilani for funding the project through Research Initiation Grant (Serial No 189).

\bmhead{Author Contribution Statement}
SD conceived and designed the analysis, performed analysis and wrote the paper.

\bmhead{Data Availability Statement}
This manuscript has no associated data or the data will not be deposited. [Authors’ comment: The datasets generated during and/or analysed during the current study are available from the corresponding author on reasonable request.]


\bibliography{sn-bibliography}


\end{document}